\documentclass[preprint]{aastex}
\usepackage{lscape} % Wes Custom
\usepackage{amsmath, amsthm,amssymb}
\usepackage{epstopdf}

\newcommand{\mHa}{\ensuremath{-1.208\pm0.004 }}
\newcommand{\mHi}{\ensuremath{-1.48\pm0.06 }}
\newcommand{\mNe}{\ensuremath{-1.4\pm0.2}}

\title{NICMOS Photometry of the Unusual Dwarf Planet Haumea and its Satellites}

\author{W. C. Fraser {1},
M. E. Brown {1},
}

\altaffiltext{1}{Geological and Planetary Sciences, California Institute of Technology}

\date{} % delete this line to display the current date

\begin{abstract}
We present here HST NICMOS F110W and F160W observations of Haumea, and its two satellites Hi'iaka and Namaka. From the measured (F110W-F160W) colours of \mHa, \mHi, and \mNe\ mag for each object, respectively, we infer that the 1.6 $\mu$m water-ice absorption feature depths on Hi'iaka and Namaka are at least as deep as that of Haumea. The light-curve of Haumea is detected in both filters, and we find that the infrared colour is bluer by $\sim 2-3\%$ at the phase of the red spot. These observations suggest that the satellites of Haumea were formed from the collision that produced the Haumea collisional family.
\end{abstract}

\begin{document}

\maketitle

\section{Introduction}

The recently named dwarf planet Haumea - formerly known as 2003 EL61 - is a peculiar Kuiper belt object; it is the primary body of a triple system \citep{Brown2006b}, the largest member of the only known Kuiper belt collisional family \citep{Brown2007}, and exhibits deep water-ice absorption which, in the Kuiper belt, has been observed only on the collisional family members, and its largest satellite - Hi'iaka \citep{Barkume2006,Schaller2008,Barkume2008}.

\begin{figure}[h] %  figure placement: here, top, bottom, or page
   \centering
   \plotone{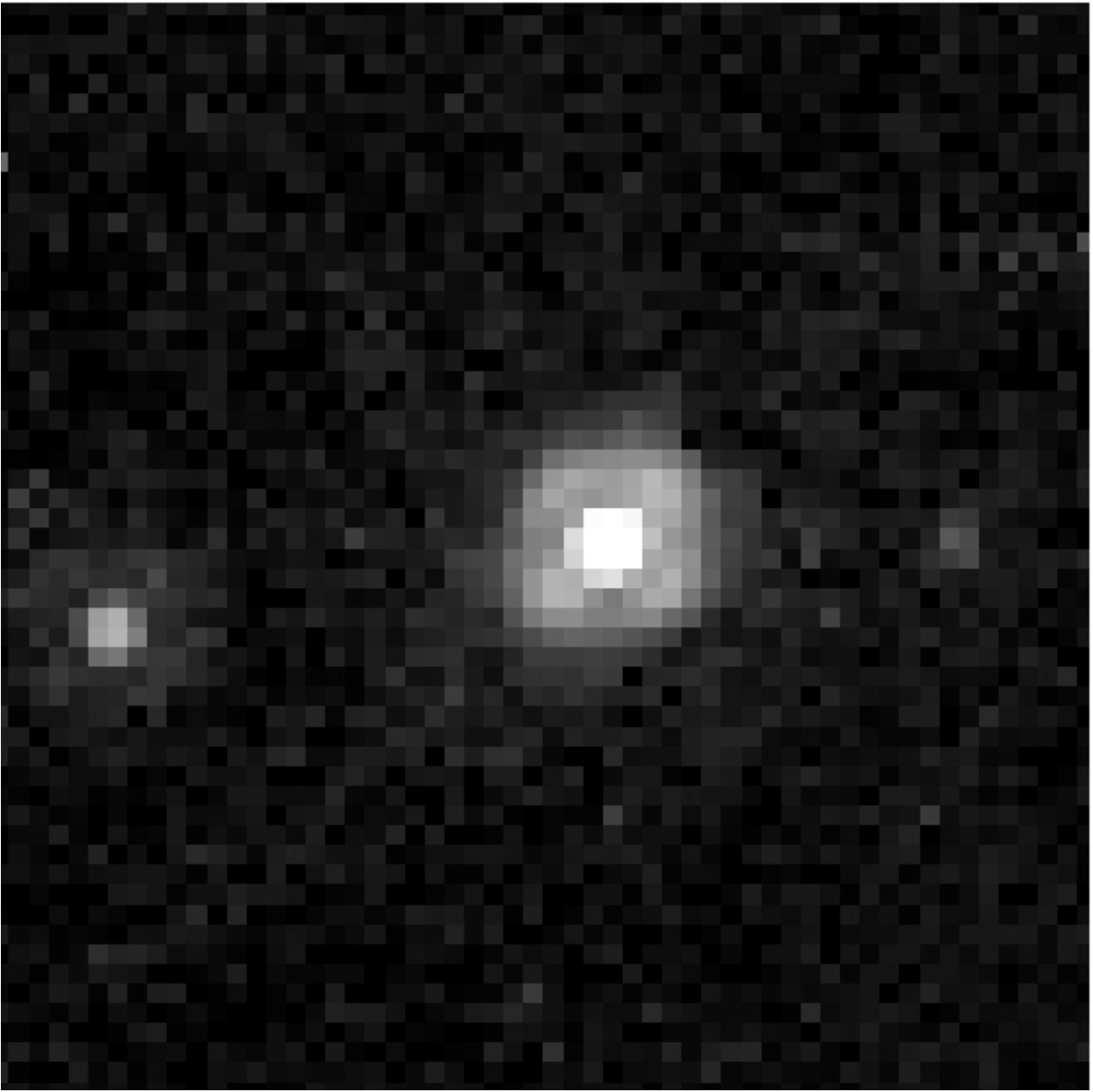} 
   \figcaption{Example F110W image of Haumea Hi'iaka (left) and Namaka (right). \label{fig:Haumea}}
\end{figure}

The small angular extent of the orbit of inner and smaller satellite, Namaka, has made a spectroscopic measure of its reflectance impossible with ground based facilities. The Hubble Space Telescope has sufficient angular resolution to separate Namaka from Haumea (see Figure~\ref{fig:Haumea}). The F160W filter is sensitive to the $\sim 1.6\mbox{ } \mu$m water-ice absorption feature. Thus with a measure of the continuum near the water-ice absorption, the depth of the water-ice absorption of Namaka can be inferred from Hubble imaging photometry.

Here we present photometry of the Haumea triple system using the NICMOS camera. With these observations, we clearly separate Haumea, Hi'iaka, and Namaka, and provide photometry of these objects with the F110W and F160W filters.
In Section 2 we describe the observations and describe data reductions we performed.
In Section 3 we present our results. We demonstrate that our measurements are consistent with a simple water-ice absorption model for Haumea, and infer the water-ice absorption of its satellites. Finally, we discuss our results in Section 4.

\section{Data and Analysis}
Observations were made in cycle 16 using NICMOS camera 1 on May 7, 2008 (UT). Images were taken in the F110W and F160W filters with exposure times of 144 and 320 s respectively. Images were taken in pairs with alternating filters, and in a five-point dither pattern to allow accurate background removal and to avoid hot-pixels in the camera. Because the time between adjacent colour pairs was short compared to the rotation period of Haumea,  the images of a pair sampled approximately the same phase of Haumea and its satellites.

The data were initially processed through {\it calnica}, the standard HST reduction and calibration procedure \citep{NICMOS1997}. Residual backgrounds and non-linearities were then removed using the {\it pedsky} and {\it rnlincor} routines from the STSDAS package\footnote{STSDAS is a product of the Space Telescope Science Institute, which is operated by AURA for NASA.}. The background of each image did not vary significantly from image to image. Thus, a median image was produced from all for each filter and removed to further subtract any residual background left over from the previous image reductions.

 We performed aperture photometry on Haumea. Aperture photometry was performed on the satellites using 5-6 pixel radius apertures, after removing the best-fit amplitude scaled {\it tinytim} PSF  \citep{Krist1993} of Haumea; we found that while the undistorted {\it tinytim} PSF could not accurately model the core of Haumea's image, subtraction of the PSF did remove the wings of Haumea's image to within the image noise near the locations of each satellite. Infinite aperture corrections were measured from the image of Haumea and the resultant aperture corrected photometry of each source is presented in Table~\ref{tab:results}. The uncertainties presented are derived from the statistical error image extensions provided with the calibrated images and do not include the $\lesssim 5\%$ absolute photometric calibration error or the $\sim 2\%$ relative calibration error from the {\it calnica} data reductions (see the NICMOS Data Reductions Handbook).

\section{Results}
\begin{figure}[h] %  figure placement: here, top, bottom, or page
   \centering
   \plotone{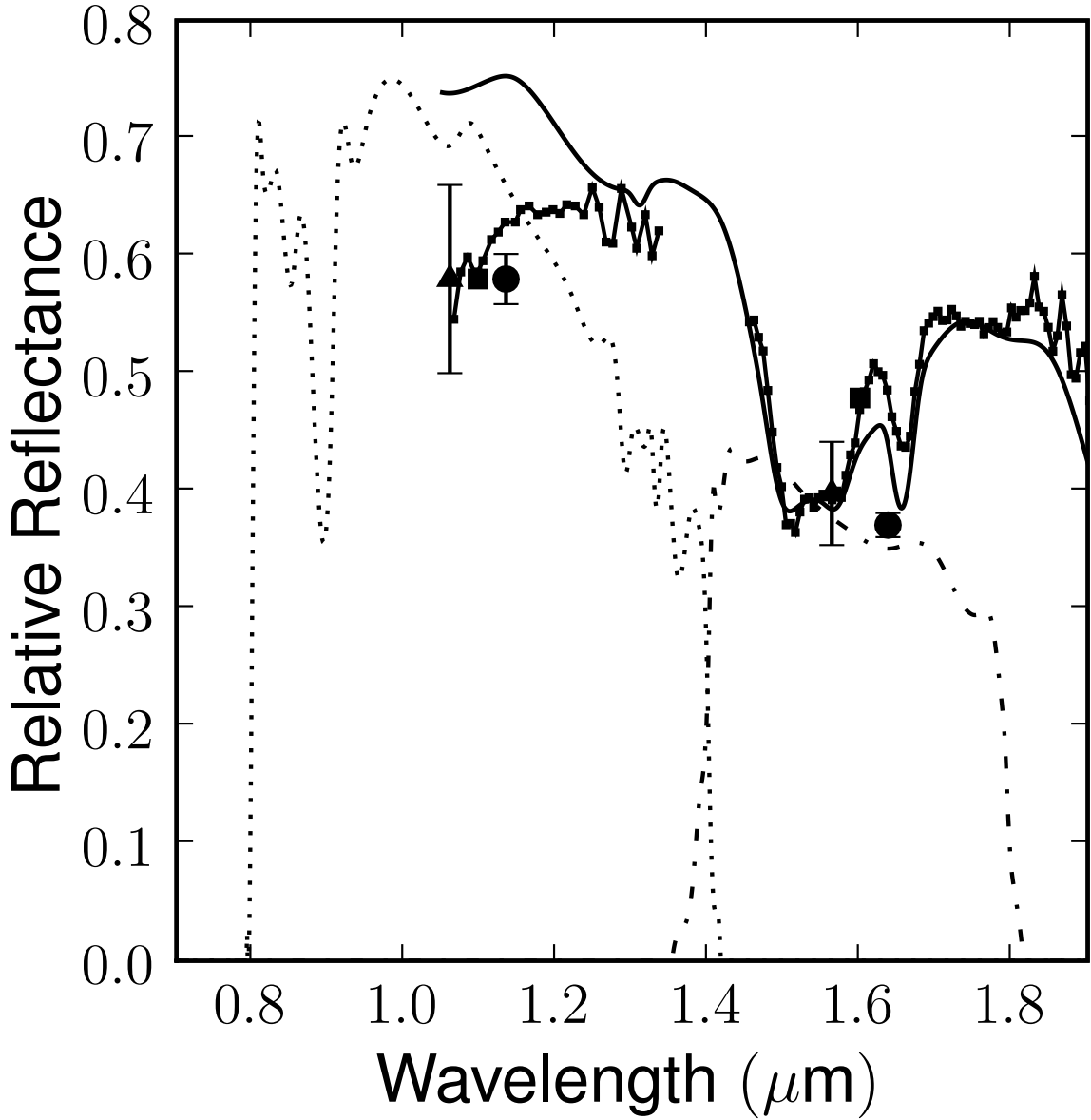} 
   \figcaption{Model water-ice plus flat component spectrum from Barkume et al. (2008) (solid line). Haumea spectrum from \citet{Trujillo2007} (dotted solid). The F110w (dotted) and F160w (dash-dotted) band-passes multiplied by the Solar spectrum are shown. Photometry of Haumea (squares), Hi'iaka (circles), and Namaka (triangles) are presented. Solar colours have been removed, and the measurements have been scaled to have equal relative flux in the F110w band. Data for Hi'iaka and Namaka have been horizontally offset for clarity. Error-bars are only the photometric shot-noise, and do not include the  relative or absolute photometric errors associated with the NICMOS filters. \label{fig:moneyShot}}
\end{figure}

 Presented in Figure~\ref{fig:moneyShot} are the average photometry of Haumea and its satellites with solar colours removed. As can be seen from this figure, all three objects exhibit absorbance consistent with water-ice.
 
 \subsection{Haumea}
We find that Haumea exhibits a mean (F110W-F160W) colour of \mHa. \citet{Barkume2008} found that the spectrum of Haumea was well described by a mixture of water-ice and a linear blue component. Using their model (Equation 1. from Barkume et al. 2008) and best-fit parameters as input to the {\it synphot} routine, we reproduce the colour of Haumea within the uncertainties of our measurements. This modeling demonstrates that from these observations, we find a $\sim 40\%$ water-ice absorption depth on Haumea compatible with the observations of \citet{Barkume2008} and can infer the relative absorptions of the two satellites.

 \citet{Rabinowitz2006} found that Haumea exhibits a double-peaked light-curve with a $\sim3.9$ hour rotation period. \citet{Lacerda2008} found that the colour of Haumea varies, and is consistent with, a large spot on one face of the dwarf planet which is optically redder than the mean surface colour.
 
 We detected Haumea's light-curve in our observations, and found that our observations are  consistent in both phase and magnitude with those of  \citet{Lacerda2008b}. The observed (F110W-F160W) exhibits a variation of $\sim0.03$ mag from the mean which is a $> 3\sigma$ deviation compared to the uncertainties in our measurements. The deviation occurs at a light-curve phase $0.78$, consistent with the phase centre of the red spot \citep{Lacerda2008}.
 
 The variation we observe is bluer, or $\sim2-3\%$ more negative $(F110W-F160W)$ at the red spot compared to the mean of the surface. The bluer apparent colour could be caused by increased water-ice absorption or larger water-ice grains compared to the mean surface. There is no reason to expect such an association of ice properties with a red visible surface component, however. A more plausible explanation is that the red spot detected by \citet{Lacerda2008} is caused by an increase in the abundance of irradiated organic materials (tholins, etc.) on the surface. These materials typically appear blue in the infrared \citep{Khare1984} and could thus account for the observed effect.
 
  %\citet{Lacerda2008b} found that Haumea's red spot exhibits an increased H-band reflectance compared to the rest of the surface. This suggests that the variation in colour we observe is caused by a variation in the water-ice absorption depth and implies a $2-3\%$ decrease of the absorption compared to the median value. This is consistent with \citet{Pinilla-Alonso2008} who found no variation in the water-ice absorption depth greater than $\sim 6\%$.

\subsection{Hi'iaka}
No photometric variation of Hi'iaka was observed to the accuracy of our measurements. Hi'iaka exhibits a deeper absoprtion than Haumea with a mean  $(F110W-F160W)=\mHi$. The observations suggest that the $\sim 1.6 \mbox{ } \mu\mbox{m}$  water-ice absorption is approximately 30-40\% deeper on Hi'iaka than Haumea which is consistent with the observations of \citet{Barkume2006}. %The mean $(F110W-F160W)$ however, cannot be reproduced with a water-ice + blue-component mixture. Further observations are required before this spectrum will be fully understood.
%This finding is consistent with \citet{Lacerda2008b} who found that the (J-H) colour of Hi'iaka is bluer than that of Haumea. 
\subsection{Namaka}

Namaka was found to exhibit a photometric variability of $\sim\pm0.3$ mag in both filters. This variability is slightly larger in amplitude than the measurement uncertainties, but further observations are required to confirm this variability. We did not detect any variability in the absorption on Namaka to within the accuracy of our measurements.  Namaka's absorption is consistent with both that of Haumea and Hi'iaka, though with larger uncertainties, and is inconsistent with a flat spectrum. The observed mean $(F110W-F160W)=\mNe$ suggests that Namaka's spectrum exhibits a water-ice absorption depth at least that of Haumea.

\section{Discussion}
Assuming similar infrared albedos, the observed mean flux ratios suggest that Haumea, Hi'iaka and Namaka have size ratios 1 : 0.29 : 0.14. The potentially large variability of Namaka suggests it has an elongated shape. This variability however, would be insignificant if the photometric uncertainty quoted here were actually slightly underestimated. Thus, more measurements are required to confirm this variability.

Our findings demonstrate that both Namaka and Hi'iaka exhibit the deep water-ice absorption characteristic of Haumea and its collisional family. If Hi'iaka and Namaka were dynamically captured satellites, they would be expected to show colours representative of the entire Kuiper belt population. They however, exhibit the deep water-ice features that are only observed on the family members.  The water-ice features exhibited by both satellites suggest that they are made of the icy-mantle material that covers the surface of Haumea \citep{Brown2007}. We conclude that the satellites are a result of the disruption that created the Haumea collisional family.

The extremely deep water-ice absorption on Hi'iaka is consistent with a blue linear-component in the water-ice mixture approximately twice as steep as that on Haumea, assuming similar mixture ratios. The deep absorption could also be caused by water-ice grains $\sim2-3$ times larger than those observed on Haumea. The true cause however, cannot be determined from our observations, and warrants further observations and spectral modeling to understand the extremely deep absorption features on this satellite.

\section{Acknowledgements}
The authors would like to thank Dr. Emily Schaller, and Darin Ragozzine for their very useful discussions of, and suggestions for this project. This material is based upon work supported by NASA under the grant NNG05GI02G. Support for program HST-GO-011169.9-A was provided by NASA through a grant from the Space Telescope Science Institute, which is operated by the Association of Universities for Research in Astronomy, Inc., under NASA contract NAS 5-26555.

%examples in case you need them
\begin{deluxetable}{lc|lll}
   \tablecaption{Photometry. All observations were made on 3:40-5:40 May 7, 2008 UT. \label{tab:results}}
  
   \startdata	
   \hline
& & \multicolumn{3}{c}{ST Magnitude} \\\hline
      UT & Filter & Haumea & Hi'iaka & Namaka \\ \hline   
	3:40 & F110W & $19.212 \pm 0.005$ & $21.91 \pm 0.02$ & $23.7 \pm 0.2$ \\
	5:14 & F110W & $19.138 \pm 0.006$ & $21.84 \pm 0.02$ & $23.3 \pm 0.1$ \\
	5:23 & F110W & $19.249 \pm 0.003$ & $21.87 \pm 0.02$ &$23.10 \pm 0.05$ \\
	5:32 & F110W & $19.260 \pm 0.005$ & $21.91 \pm 0.08$ &$23.45 \pm 0.05$ \\
	5:41 & F110W & $19.32 \pm 0.01$ 	  & $21.85 \pm 0.05$ & $23.8 \pm 0.3$ \\ \hline
	3:42 & F160W & $20.422 \pm 0.005$ & $23.41 \pm 0.01$ & $25.0 \pm 0.1$ \\
	5:17 & F160W & $20.319 \pm 0.006$ & $23.38 \pm 0.03$ & $24.6 \pm 0.1$ \\
	5:26 & F160W & $20.454 \pm 0.004$ & $23.44 \pm 0.02$ &$24.5 \pm 0.1$ \\
	5:35 & F160W & $20.490 \pm 0.005$ & $23.33 \pm 0.09$ &$24.9 \pm 0.1$ \\
	5:44 & F160W & $20.53 \pm 0.01$ 	  & $23.25 \pm 0.04$ & $25.4 \pm 0.3$ \\
				
   \enddata
\end{deluxetable}
   
\begin{deluxetable}{c|l|l|l|}
   \tablecaption{Colour measurements. All measurements do not include the 2\% relative or $\lesssim 5\%$ absolute photometric error associated with the NICMOS filters. This uncertainties however, will cause a common offset in the colours for each measurement and the relative differences between measurements will remain constant. Numbers in parenthesis are the rotation phase of Haumea determined from the published light-curve of Lacerda et al. (2008).\label{tab:colours}}
   \startdata	
   \hline
     & \multicolumn{3}{|c}{(F110W-F160W) (ST Magnitude)} \\ \hline
      UT &  Haumea & Hi'iaka & Namaka \\ \hline
	3:40 &  $-1.209 \pm 0.007 (0.36)$ & $-1.50 \pm 0.04$ & $-1.25 \pm 0.2$ \\
	5:14 &  $-1.182 \pm 0.008$ (0.78) & $-1.54 \pm 0.04$ & $-1.3 \pm 0.2$ \\
	5:23 &  $-1.205 \pm 0.005$ (0.82) & $-1.57 \pm 0.04$ &$-1.44 \pm 0.05$ \\
	5:32 &  $-1.230 \pm 0.007$ (0.86) & $-1.4 \pm 0.1$ &$-1.42 \pm 0.05$ \\
	5:41 &  $-1.22 \pm 0.01$ (0.89) & $-1.40 \pm 0.07$ & $-1.6 \pm 0.4$ \\
				
   \enddata
\end{deluxetable}

\end{document}